\documentclass[conference]{IEEEtran}
\IEEEoverridecommandlockouts
\newcommand{\reals}{{\mbox{\bf R}}}

\newcommand{\diag}{\mathop{\bf diag}}

\newcommand{\eg}{{\it e.g.}}
\newcommand{\ie}{{\it i.e.}}

\newcommand{\BEAS}{\begin{eqnarray*}}
\newcommand{\EEAS}{\end{eqnarray*}}
\newcommand{\BEA}{\begin{eqnarray}}
\newcommand{\EEA}{\end{eqnarray}}
\newcommand{\BEQ}{\begin{equation}}
\newcommand{\EEQ}{\end{equation}}
\newcommand{\BIT}{\begin{itemize}}
\newcommand{\EIT}{\end{itemize}}

\newcounter{algorithmctr}[section]
\renewcommand{\thealgorithmctr}{\thesection.\arabic{algorithmctr}}
   {\refstepcounter{algorithmctr}\begin{list}{}{%
       \setlength{\rightmargin}{0\linewidth}%
       \setlength{\leftmargin}{.05\linewidth}}%
       \rmfamily\small
       \item[]{\setlength{\parskip}{0ex}\hrulefill\par%
        \nopagebreak{\bfseries\textsf{Algorithm \thealgorithmctr~}}}}%
   {{\setlength{\parskip}{-1ex}\nopagebreak\par\hrulefill} \end{list}}

\usepackage{cite}
\usepackage{amsmath,amssymb,amsfonts,siunitx}
\usepackage{algorithmic}
\usepackage{graphicx}
\usepackage{textcomp}
\usepackage{xcolor}
\usepackage{pgf,pgfplots}
\usepackage{import}
\usepackage{booktabs}
\def\BibTeX{{\rm B\kern-.05em{\sc i\kern-.025em b}\kern-.08em
    T\kern-.1667em\lower.7ex\hbox{E}\kern-.125emX}}
\begin{document}

\title{Estimation of Shade Losses in Unlabeled PV Data
\thanks{This material is based on work supported by the U.S. Department of 
Energy's Office of Energy Efficiency and Renewable Energy (EERE) under the 
Solar Energy Technologies Award Number 38529.}
}

\author{\IEEEauthorblockN{Bennet Meyers$^{1,2}$ and David Jose Florez 
Rodriguez$^2$}
	\IEEEauthorblockA{$^1$ SLAC National Accelerator Laboratory, Menlo Park, 
	CA, 94025, USA \\$^2$ Stanford University, Stanford, CA, 94305, USA}}

\maketitle

\begin{abstract}
We provide a methodology for estimating the losses due to shade in power 
generation data sets produced by real-world
photovoltaic (PV) systems. We focus this work on estimating shade loss from 
data that are unlabeled, \ie~power 
measurements with time stamps but no other information such as site 
configuration or meteorological data. This approach enables, for the first 
time, the analysis of data generated by small scale, distributed PV systems, 
which do not have the data quality or richness of large, utility-scale PV 
systems or research-grade installations. This work is an application of the 
newly published signal decomposition (SD) framework, which provides an 
extensible approach for estimating hidden components in time-series data.
\end{abstract}

\begin{IEEEkeywords}
photovoltaic systems, solar energy, distributed power generation, energy 
informatics, signal processing, machine learning, statistical learning, 
unsupervised learning
\end{IEEEkeywords}

\section{Introduction}

The distributed rooftop solar market is growing 
rapidly, with 3.2 GW$_\text{DC}$ of residential PV installed in the U.S. in 
2020, the 
largest year on 
record~\cite{SEIA2021}. While utility 
solar continues to contribute the majority of new installations, 46\% of new 
installed capacity in 2020 
(5.3 GWdc) were non-utility, distributed systems~\cite{SEIA2021}. Clearly, 
distributed PV systems continue to account for a significant fraction of solar 
power generation.

However, distributed PV presents a distinct challenge for data analysis, as 
compared to utility PV plants. We call the data that come from 
utility power plants \emph{labeled}. Generally speaking, this means the power 
production data comes with supplementary data and information such as 
correlated measurements from 
meteorological stations and system configuration information from engineering 
drawings. In contrast, distributed PV tends to generate \emph{unlabeled} and 
\emph{partially labeled} data sets, making it difficult or impossible to 
calculate a performance index for the systems (see~\cite{Townsend1994} for 
definition and description of performance index in this context). 

In this manuscript, we present and validate a novel approach for estimating the 
shade losses in PV systems from unlabeled production data, \ie, the 
measured power output of the system over a multi-year period. We assume that 
the power measurements are taken on regular intervals on a sub-daily basis, 
typically in the range of every 5 minutes to once an hour. The method takes the 
unlabeled data for a single PV system as an input, and returns an estimate of 
the energy lost to shade. Our approach is an application of the \emph{signal 
decomposition (SD) 
framework}~\cite{Meyers2022}, in which we model 
the shade loss as one of a number of component processes which combine to 
generate the observed signal. The methods proposed in this paper are available 
as a module in the Solar Data Tools package~\cite{Meyers2020b,solar-data-tools}.
% This manuscript should be 
%considered a companion to ``Estimation of Soiling Losses in Unlabeled PV 
%Data,'' which is also based on the SD framework, also submitted to this 
%conference~\cite{pvsc-soiling}.

\section{Background and related work}\label{s-background}

The non-linear loss of power generation in PV systems experiencing partial 
shade has long been known to be a major concern in real world systems, 
particularly for distributed PV in urban 
environments~\cite{Woyte2003,Deline2009,Meyers2017}. Generally speaking, prior 
research into quantifying shade losses in operational PV systems has focused on 
generating a high quality model of the system in question to use as a 
reference~\cite{MacAlpine2017,Torres18,Fairbrother2021} or utilizing additional 
data sources, such as an unshaded reference system~\cite{Tsafarakis2019} or 
current-voltage response measurements (\ie, IV 
curves)~\cite{Liu2019}.

In this paper, we develop a methodology for estimating the soiling losses in 
unlabeled PV system power generation data, \ie, a time series of real power 
measurements and nothing else. This approach, while admittedly less accurate 
than some of the methods mentioned above, unlocks large swaths of real-world 
production data that otherwise could not be analyzed for shade loss. 
Additionally, the analysis is easily automated, enabling the analysis of large 
fleets of heterogeneous PV systems. This approach is based on a framework of 
signal 
decomposition (SD), and a summary of the long history of signal decomposition 
and many related topics involved is available in~\cite[\S3]{Meyers2022}.

\section{Methods}

We construct an SD problem~\cite{Meyers2022} that models the decomposition of 
the power data into components that represent weather effects, a clear sky base 
line, and shade losses. This approach can be thought of as 
an \emph{unsupervised} machine learning (ML) method for finding structure in 
time series data, similar to model-based clustering methods like Gaussian 
mixture models~\cite[\S14.3.7]{Hastie2009}. Unlike supervised ML, there is no 
``training'' of the method; we simply design the mathematical optimization 
problem, input the data for analysis, and receive the estimate of the shade 
losses. 

As important as the formulation of the SD problem, however, is how the raw 
power data is prepared for analysis. Before estimating the components, the 
power data is filtered and transformed in a way that emphasizes the periodic 
nature of fixed shade patterns, both day-to-day as well as season-to-season and 
year-to-year. We find that putting the data in a particular form is critical to 
being able to apply the SD framework to this problem.

In the remainder of this section, we describe the data preparation, SD 
formulation, 
and validation procedure.

\subsection{Data preparation}

\begin{figure}
\centering
\resizebox{\columnwidth}{!}{
\import{figs/}{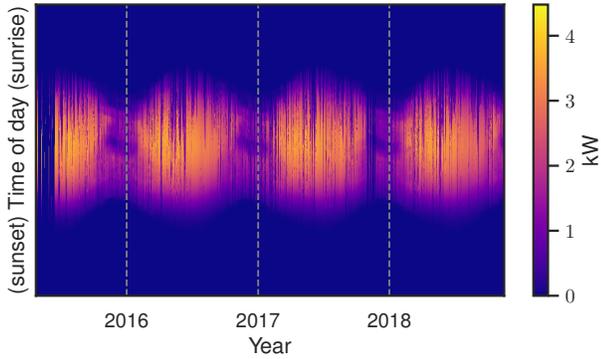}
}
\caption{An example multi-year, unlabeled power data set, with noticeable shade 
in the winter.}
\label{f-data}
\end{figure}

We begin with a multi-year data set of measured power from a PV system, that 
may or may not have shade. A typical example is shown in figure~\ref{f-data}, 
with the 5-minute real power measurements represented as a heatmap with lighter 
pixels representing higher power. In this view, days are columns of pixels, 
starting with sunrise on the top of the image and ending with sunset at the 
bottom. Consecutive days go left to right. The seasons are visible as a 
shortening and lengthening of the day, seen as width of the bright portion. 
This example happens to have noticeable shade in the winter months, observable 
as a dark patch in the middle of the shorter days. We can already observe that 
this pattern repeats itself on a yearly basis and has a different character 
than power lost to clouds, which are seen as dark, vertical lines.

\begin{figure}
\centering
\resizebox{\columnwidth}{!}{
\import{figs/}{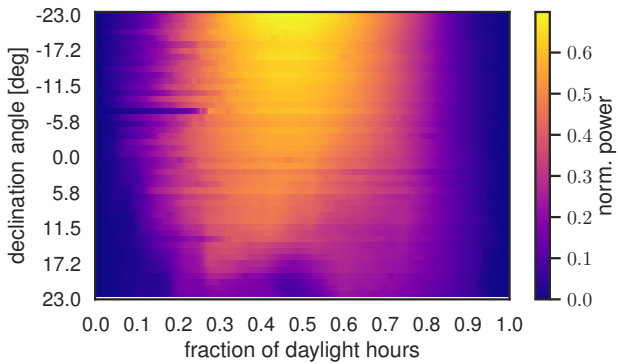}
}
\caption{The output of the data preparation procedure for the data shown in 
figure~\ref{f-data}. We call this the ``transformed'' data.}
\label{f-proc}
\end{figure}

The end result of the data preparation process for this example is shown in 
figure~\ref{f-proc}. In the remainder of this section, we will describe how 
this second figure is generated.

\paragraph{Onboarding}
First, the data is onboarded and standardized using Solar 
Data Tools (SDT)~\cite{Meyers2020b,solar-data-tools}. This accounts for gaps in 
data or missing time stamps and puts the data in the 2D array form shown in 
figure~\ref{f-data}. Gaps in the original data are either linearly interpolated 
or zero-filled depending on if it is day or night. The software has 
tools for correcting common errors in PV data sets such as time shifts due to 
daylight savings clock changes. 

\paragraph{SDT data labeling}
With the original scalar time series now embedded in a matrix, we use two 
additional features of SDT: clear day detection and sunrise/sunset detection. 
The first subroutine automatically segments the data by categorizing days as 
being (more) sunny or (more) cloudy. The second subroutine analyzes the matrix 
embedding of the power signal to make a robust estimate of local sunrise and 
sunset times each day. (These may be calculated exactly using well known 
equations if the latitude and longitude of the PV systems is known, but we 
assume \emph{no} access to outside information besides the measured power.) We 
note briefly here that both these subroutines in SDT make use of the SD 
framework~\cite{Meyers2022} and operate on the general principle of detecting a 
structured component within a signal.

\paragraph{Soiling estimation and correction}
Next, we apply a method for detecting soiling trends in unlabeled PV power data 
sets, which is described in soon to be published manuscript~\cite{pvsc-soiling} 
and also makes use of an SD framework. As described 
in~\cite[\S3-D]{pvsc-soiling}, we use the estimated soiling component to 
correct the measured data, generating a new estimate of system power, with the 
soiling losses simply removed. This procedure is analogous to generating a 
performance index that accounts for estimated soiling, but it does not require 
any additional information besides power to construct. We find 
that removing detected soiling trends in the data yields a significant 
improvement in the subsequent shade analysis.

\begin{figure}
\centering
\resizebox{\columnwidth}{!}{
\import{figs/}{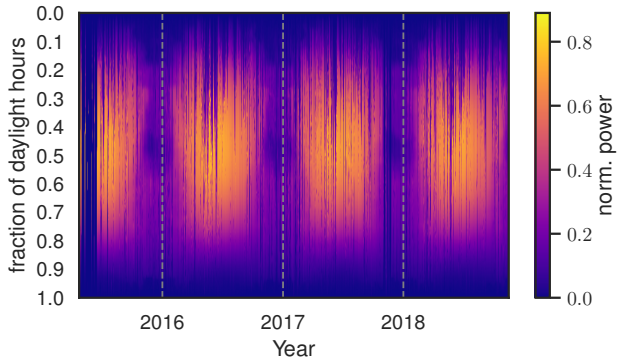}
}
\caption{The result of the masking and resampling step in the data preparation 
procedure.}
\label{f-normed}
\end{figure}

\paragraph{Masking and resampling}
Using the estimates of sunrise and sunset times, a mask is generated to remove 
the night time data. This results in a different number of data points each 
day, as the length of the day changes, so the resulting signals are upsampled 
using linear interpolation to all have the same number of samples per day, 
chosen to be 256. (This value is selected somewhat arbitrarily to be a power of 
2 larger than any of the resulting day-long signal segments.) Additionally at 
this step, the data is normalized to be in the 
range $[0,1]$. This results in a transformation of the original power matrix 
shown in figure~\ref{f-normed}. We observe that the days have been 
``stretched'' so that they all appear to be the same length.

\paragraph{Averaging by declination angle bin}
The final step is to average the data across seasons and years, excluding the 
cloudy days detected with SDT. We do this by assigning a \emph{declination 
angle} to each day, which is the angular position of the sun at solar noon, 
with respect to the plane of the equator, and is given by a simple equation 
based only on the day of year~\cite[\S1.6]{duffie_beckman_2013}. We 
then bin 
the declination angle values, and keeping only the (more) sunny days, we 
average the daily signals in each bin. Finally, we arrange the bin in 
sequential order, producing the tabular data set shown in figure~\ref{f-proc}. 
If for whatever reason there are no data available to average for a particular 
bin (\eg, due to missing data and bad weather, no clear sky data was recorded 
at that declination angle), we simply fill that row with missing values 
markers, or `$?$' in the notation of SD~\cite{Meyers2022}.

We treat this tabular data 
representation as the vector-valued input $y$ to the SD problem. Using the 
notation from~\cite{Meyers2022}), this gives us dimensions for our SD problem 
of $T=47$ and $p=256$.

\subsection{SD problem formulation}

Utilizing the notation of signal decomposition defined in~\cite{Meyers2022}, we 
say that our data for this SD problem instance is a signal $y\in\left(\reals 
\cup \{?\} \right)^{T\times p}$ with length 
$T=47$ is the number of declination angle bins and $p=256$ is the number of 
upsampled normalized power measurements in one day. In other words, our 
signal, $y$, is a matrix of size 47 by 256, possibly with missing values. We 
model the signal $y$ as the 
composition of $K=3$ components, $x^1$ to $x^3$, the sum of which must be equal 
to the signal $y$ at the entries that do not contain missing values, or in 
other words,
\[y_{t,i}  = x_{t,i}^1 + x_{t,i}^2 + x_{t,i}^3,\mbox{ for }t\in\mathcal{K},\]
where $\mathcal{K}$ is the set of indices $(t,i)$ that do not contain missing 
values (\ie, the ``known'' set). The three 
components are defined in the SD model by their cost functions $\phi_k(x^k)$ 
for 
$k=1,\ldots,3$. (We drop the superscript $k$ on $x$ to keep the notation 
lighter when not distinguishing between particular components.) Following the 
convention in~\cite{Meyers2022}, we assign the first component $x_1$ the role 
of the ``residual'' and the last component $x_3$ to be what we are 
interested in calculating, in this case shade loss. (We think of this ordering 
as ``building'' the SD model, by accounting for sourcing of variation in the 
data before arriving at the component of interest.) Below, we describe 
each of these components and their mathematical definition within the SD 
framework. 

\paragraph{Component $x^1$}
The first component represents 
the residual of the model, and it represents the remaining variability due to 
weather that was not removed through the clear day filtering with SDT. It is 
taken to be the sum-absolute cost 
function~\cite[\S6.1]{convex_opt},
\BEQ
\phi_1(x) = \sum_{(t,i)} \left|x_{t,i}\right|,
\EEQ
which is chosen for its robustness to outliers and preference for values that 
are exactly zero.

\paragraph{Component $x^2$}
We construct the second component to represent the 
unobstructed clear sky behavior of a fixed-tilt PV system. We define 
$\phi_2(x)$ to represent the distance from a corpus of clear sky signals for 
all 
possible tilts and azimuth, generated by pvlib-python~\cite{Holmgren2018}, and 
then normalized and transformed the same as the measured data. We 
follow the methodology described in~\cite[\S6.3]{Meyers2022} for fitting a 
component class loss to a corpus of signals known to belong to the class. This 
is done by first calculating the statistical mean $\mu$ and covariance 
$\Sigma$ of the signal corpus, arranged as a matrix. Then, we generate a 
low-rank approximation calculating the eigendecomposition~\cite[\S8]{Golub1996} 
of the 
empirical covariance,
\[\Sigma = Q\Lambda Q^{-1},\]
and keeping the top $k=6$ eigenvalues, giving us $Q_k\in\reals^{p \times k}$ 
and $\Lambda_k=\diag(\lambda_k)$ with $\lambda_k\in\reals^k$. We then construct 
the component cost as the linear combination of three consituent costs
\BEQ
\phi_2(x) = \lambda_{2a} \ell_{2a}(x) + \lambda_{2b}(x) \ell_{2b} + I_2(x),
\EEQ
where $\ell_{2a},\ell_{2b}:\reals^{T\times p}\rightarrow \reals$ are continuous 
functions, $\lambda_{2a}, \lambda_{2b}$ are scalar problem weights, and 
$I_2:\reals^{T\times p}\rightarrow \{0, \infty\}$ is an 
indicator 
function of a set of equality and inequality constraints imposed on the second 
component. The first loss applies regularization to a helper variable which 
will indicate how much of which corpus eigenvectors to use,
\[\ell_{2a}(x) = \|M z_{2}\|_F,\]
with $z_{2a}\in\reals^{k\times p}$ is the new variable in the SD optimization 
problem and $M\in\reals^{T\times k}$. The function $\|\cdot\|_F$ is the 
standard Frobenius matrix norm~\cite[\S A.1.1]{convex_opt}. The matrix $M$ is 
calculated from the 
reduced 
eigenvalues calculated earlier,
\[M = \left[\begin{matrix}
\diag(1/ \lambda_k) \\ \mathbf{0}
\end{matrix}\right].\]
The second loss selects for components that change smoothly with changing 
declination angle,
\[\ell_{2b}(x) = \|D_2 x\|_F,\]
where $D_2\in\reals^{(T-2)\times T}$ is the second-order discrete difference 
operator. (See, for example, \cite[\S6.4]{Boyd2018} for information on 
difference matrices.) In other words, these first two terms tell us that we 
want signals that are similar to our corpus, and changing smoothly day-to-day. 
Finally, the constraints in $I_2$ are the union of the following
\begin{eqnarray*}
\{x &\mid& x \succeq 0\} \\
\{x &\mid& D_2 x \preceq 0\} \\
\{x &\mid& x_{t, i} = 0\mbox{ for all }t\mbox{ and }i\in\{1, 256\}\} \\
\{x &\mid& x - \mu = z_2^T Q_r^T \}.
\end{eqnarray*}
Briefly, this constrains the component to (1) be non-negative, (2) have 
non-positive curvature along the the first axis (declination angle), (3) start 
and end at zero along the second axis (fraction of day), and (4) be a 
combination of corpus eigenvectors. Taken altogether, we have constructed a 
component class $\phi_2$ that is bespoke for this particular analysis and 
provides a baseline to reference the losses due to shade in the data.
 
\paragraph{Shade component}
The third component represents the shade 
losses in the system. We define it mathematically as
\[
\phi_3(x) = \lambda_3 \left(\|D_2 x\|_F + \|D_2 x^T\|_F\right) + I_{-}(x),
\]
where $\lambda_3$ is a problem parameter and $I_{-}$ is the indicator function 
of the nonpositive orthant. This cost encourages component values that 
do not fluctuate wildly along either axis (declination angle or day fraction) 
and are nonpositive (shade can only reduce the power production). A key insight 
of our data preparation process is that in the transformed data space, periods 
of shade tend to be ``compact'', \ie, well localized in space with clear 
boundries.

\subsection{SD parameters}
The SD problem formulation includes three parameters, 
$\lambda_{2a}$, $\lambda_{2b}$, and $\lambda_3$. These parameters are 
tunable, and different values can greatly affect the characteristics and 
quality of the resulting decomposition. A deep discussion on the role of 
parameters in SD problems can be found in~\cite[\S2.6]{Meyers2022}. While a 
method is provided in~\cite[\S2.7]{Meyers2022} for selecting optimal parameter 
values, we find that in this context it makes more sense to rely on the 
practical experience of the analyst. In other words, we have found values 
that work well in many cases, and a small amount of hand-tuning is accepted in 
other cases. A description of these parameters and their default values are 
given in table~\ref{t-params}.

\begin{table}
\centering
\caption{SD problem parameters}
\begin{tabular}{ccp{5cm}}
\toprule
param. & value & description \\\midrule
$\lambda_{2a}$ & 0.5& penalizes distance of clear sky component from corpus of 
reference signals \\
$\lambda_{2b}$ & \num{1e3} & encourages smoothness of clear sky component along 
declination angle axis \\
$\lambda_3$ & 1 & encourages compact shade component  
\\\bottomrule
\end{tabular}
\label{t-params}
\end{table}

\subsection{Validation}\label{s-validation}

As described in detail in~\cite{pvsc-shade-david}, a small fleet of 25 rooftop 
PV 
installations in Southern California has been modeled in 
pvlib-python~\cite{Holmgren2018}. We briefly describe the procedure here but 
refer readers to that paper for a complete description of the modeling building 
process. 

We estimate the unobstructed power 
output of the system over a typical clear sky year, making use of the 
simplified Solis clear sky model~\cite{solis} available in 
pvlib-python~\cite{Holmgren2018}. This provides us 
with a reference estimate for the unshaded power production of these systems, 
which we can use to validate the estimate from the SD formulation. 

In this paper, we validate the proposed method for shade loss estimation based 
on signal decomposition by using the pvlib models of the systems as a 
reference. An estimate of the shade loss under clear sky conditions is 
generated for each site by comparing the sunny days from the data set to the 
pvlib model. This calculation is taken as the ground truth, and we then 
calculate errors between the SD estimate and the pvlib estimate.

\section{Results}

In this section, we present a validation of the proposed SD 
methodology for shade loss estimation. We compare the estimates of shade loss 
generated by the SD methodology to the estimates derived from modeling the 
sites in pvlib-python, as described in~\cite{pvsc-shade-david}. First, we 
discuss in more detail a case 
study of a single site. Then we discuss the analysis of the fleet. 

We find that 
the proposed method clearly detects losses in the systems that are consistent 
with the pvlib estimates of shade loss. For low-to-medium shaded systems that 
have azimuth orientations within $90^\circ$ of due south, the method does quite 
well in capturing the temporal impacts of shade and estimating the total 
losses. 
As the shade losses get more extreme, the SD methodology tends to 
conservatively under-predict the total losses. Conversely, we find that the SD 
methodology tends to \emph{over}-predict shade losses for systems with azimuths 
pointing due east and west. Improving the performance of the algorithm systems 
with these azimuths will be an area of future research.

\subsection{Case study}

\begin{figure}
\centering
\resizebox{\columnwidth}{!}{
\import{figs/}{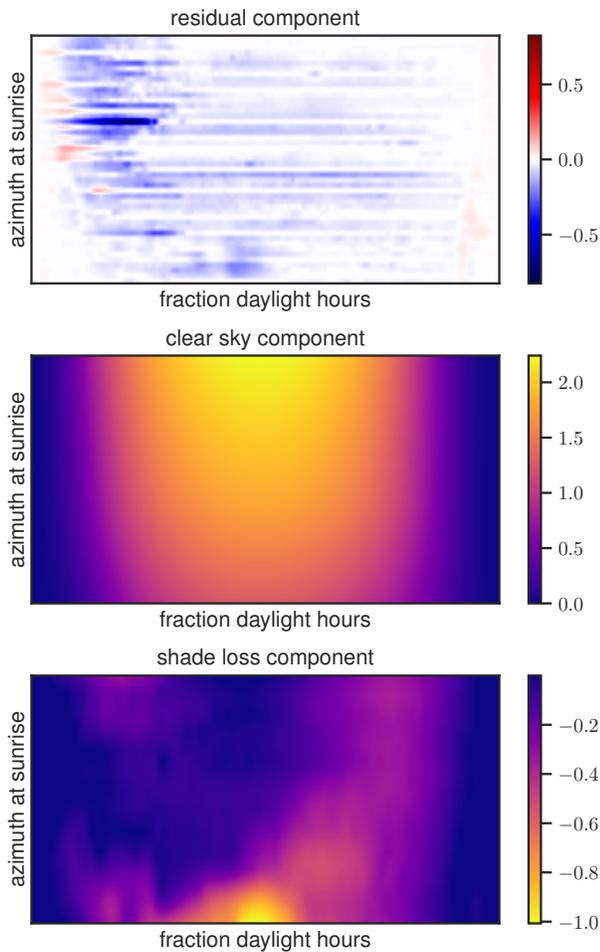}
}
\caption{The calculated decomposition of the signal shown in 
figure~\ref{f-proc}. Top, the weather component. Middle, the clear sky 
component. Bottom, the shade loss component.}
\label{f-decompose-example}
\end{figure}

A typical example data set from the validation fleet of 25 systems is shown 
in figure~\ref{f-data}. The resulting decomposition for this data set is shown 
in figure~\ref{f-decompose-example}, with the components displayed in the 
transformed data space. 

\begin{figure}
\centering
\resizebox{\columnwidth}{!}{
\import{figs/}{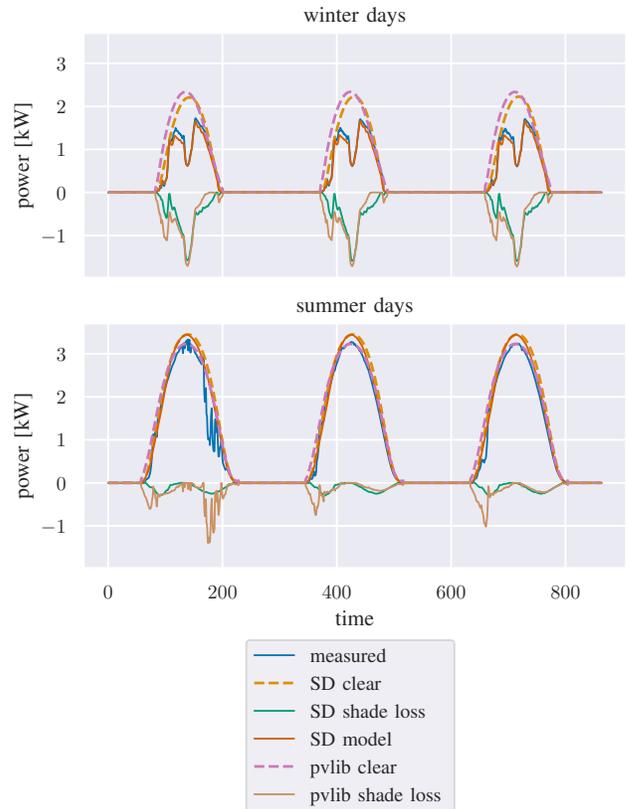}
}
\caption{The calculated decomposition of the signal shown in 
figure~\ref{f-proc}, shown as a time series plot.}
\label{f-decompose-example-ts}
\end{figure}

We can view the decomposition in the original time 
series space by undoing the data transformation process. A selection of days 
and their SD decompositions are shown in figure~\ref{f-decompose-example-ts}, 
along with the pvlib estimates of clear sky power output and soiling loss. We 
see good general agreement, but we note that the SD model is tending to 
underpredict the clear sky component under highly shaded conditions, as seen in 
the winter. This is reasonable given our model formulation and lack of any 
other reference. We generally prefer 
for the smallest reasonable decomposition that explains the data. (Note that 
the loss can be made arbitrarily large by increasing the magnitude of the clear 
sky component.)

%By reversing the data preprocessing transformation, we can ``unroll'' the 
%shade 
%loss term, and label the original input data set to show periods of high 
%shading, as illustrated in figure~\ref{f-shade-annot}. In this case, we apply 
%a 
%threshold of highlight losses of greater than 50\% of the unshaded power. 

\begin{figure}
\centering
\resizebox{\columnwidth}{!}{
\import{figs/}{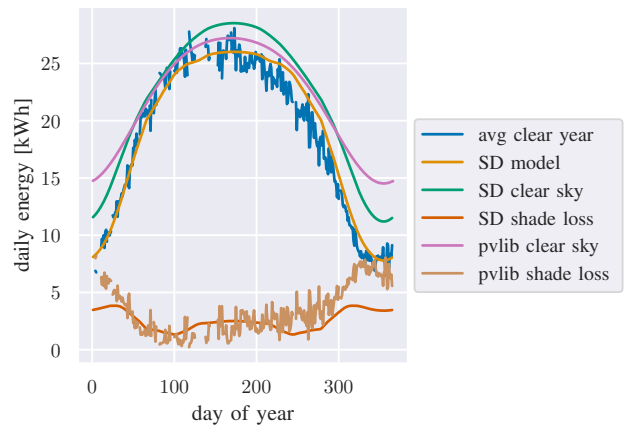}
}
\caption{Analysis of the shade OSD components that estimates the seasonal 
energy lost to shade.}
\label{f-shade-energy}
\end{figure}

An analysis of the shade losses in a typical year for this data set is shown in 
figure~\ref{f-shade-energy}. We see, again, that the SD model tends to be 
conservative; the proposed algorithm estimates a lower clear sky baseline in 
the winter than the pvlib estimate, resulting in smaller estimated shade losses 
in the winter. However, the SD estimate of shade losses is of the correct 
magnitude and has the same seasonal structure as what we estimate using pvlib.

We calculate two summary error metrics from this yearly analysis: (1) 
the shade loss root-mean-square error (RMSE) and (2) the difference between 
pv-lib and SD estimates of total yearly shade loss as a percentage of total 
yearly energy (RE, for ``relative error''). For this example, the RMSE is 1.61 
kWh and the difference 
in 
energy loss is 2.8\%.

%\begin{figure}
%\centering
%\resizebox{\columnwidth}{!}{
%\import{figs/}{shade-annot.pgf}
%}
%\caption{A heatmap of PV power annotated to mark periods of time that are 
%represent more than 5 0\% loss of power due to shade.}
%\label{f-shade-annot}
%\end{figure}

\subsection{Fleet study}

\begin{figure}
\centering
\resizebox{\columnwidth}{!}{
\import{figs/}{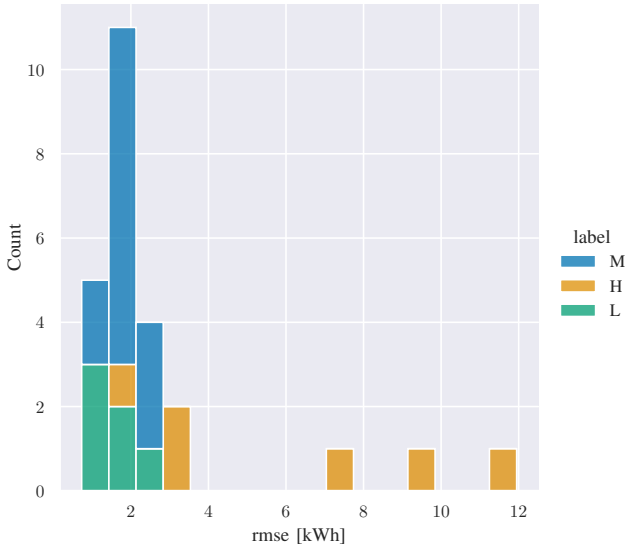}
}
\caption{Distribution of RMSE of seasonal shade loss across the fleet. The 
colors show the low, medium, and high shade bins.}
\label{f-fleet-rmse}
\end{figure}

\begin{figure}
\centering
\resizebox{\columnwidth}{!}{
\import{figs/}{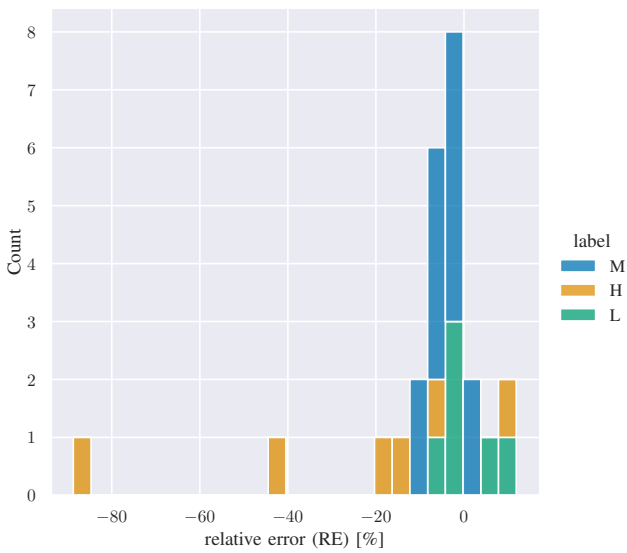}
}
\caption{Distribution of relative error (RE) across the fleet, which is 
defined 
as the difference between the SD model yearly loss estimate and the pvlib 
yearly loss estimate, normalized for total energy capture. The colors show the 
low, medium, and high shade bins.}
\label{f-fleet-re}
\end{figure}

\begin{figure}
\centering
\resizebox{\columnwidth}{!}{
\import{figs/}{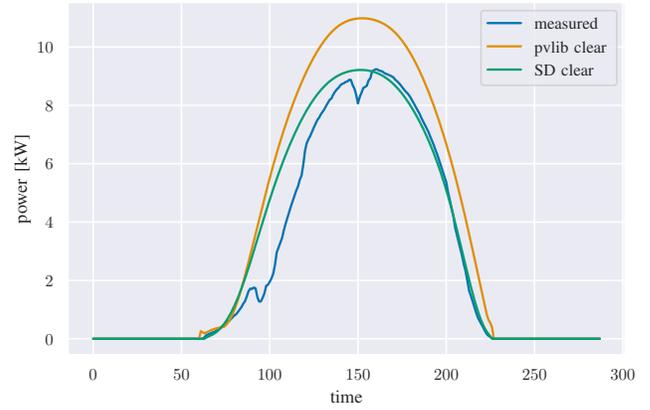}
}
\caption{A view of a single day of data (measured) and the pvlib and SD 
estimates of the clear sky envelope. This data was selected from the system in 
the high shade group with the worst error metrics. Note that the green line is 
below the orange line.}
\label{f-high-shade-example}
\end{figure}

\begin{figure}
\centering
\resizebox{\columnwidth}{!}{
\import{figs/}{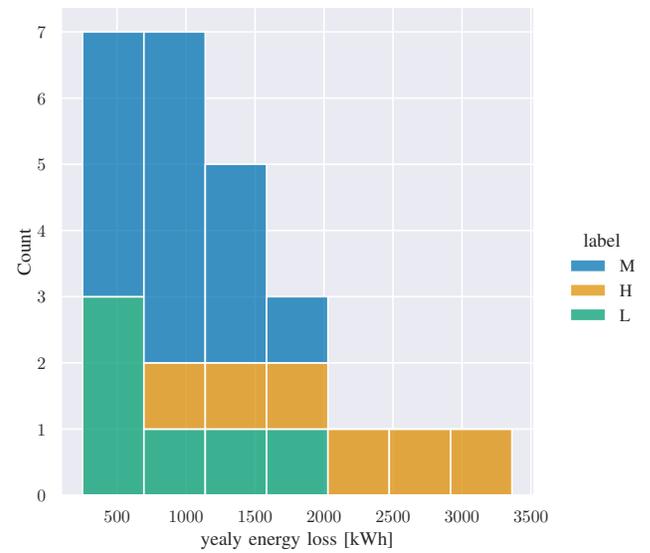}
}
\caption{Distribution of SD estimates total yearly energy lost to shading. As 
expected, the high shade sites have the largest total shade losses.}
\label{f-fleet-total-losses}
\end{figure}

We present the RMSE and RE for the 25 sites in the validation fleet in 
figures~\ref{f-fleet-rmse} and~\ref{f-fleet-re} respectively.  We hand labeled 
each of the 25 sites as one of three bins: low (L), medium (M), and high 
(H) shade losses. The mean and standard deviation of these metrics are given 
for 
each bin in table~\ref{t-error-summary} We observe that the high-shade bin 
include sites that are the largest outliers, and that those high error sites 
are negative, \ie, the SD method under-predicts 
the amount of shade. Additionally, this effect gets larger with more 
aggressive 
shade. An extreme example is shown in figure~\ref{f-high-shade-example}. 
Visual inspection of this site confirms trees directly to the south of the 
system, which project shade all day long. As expected, SD estimates a smaller 
clear sky envelope for the data, evidently the smallest such envelop that is 
consistent with the data. However, we note that the SD model nonetheless 
estimates large yearly total shade 
losses for highly shaded sites, 
as shown in figure~\ref{f-fleet-total-losses}. As discussed with the case 
study, the SD model can be 
thought of as a lower bound on the shade losses.

\begin{table}
\centering
\caption{Summary of SD errors}
\begin{tabular}{ccccc}
\toprule
& \multicolumn{2}{c}{RMSE} & \multicolumn{2}{c}{RE} \\ 
bin & mean & std & mean & std \\\midrule
L & 1.63 & 0.60 & 0.65 &  6.26 \\
M & 1.83 & 0.52 & -3.94 & 3.30 \\
H & 6.22 & 3.95 & -26.1 & 35.3 \\\bottomrule
\end{tabular}
\label{t-error-summary}
\end{table}

\subsection{Azimuth dependence}
\begin{figure}
\centering
\resizebox{\columnwidth}{!}{
\import{figs/}{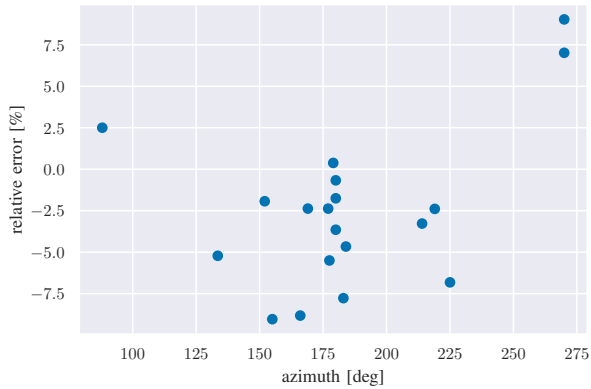}
}
\caption{The dependency of relative error on azimuth. The azimuth angle is 
defined with $0^\circ$ being north and the angle increasing in the 
counterclockwise direction.}
\label{f-azimuth}
\end{figure}

\begin{figure}
\centering
\resizebox{\columnwidth}{!}{
\import{figs/}{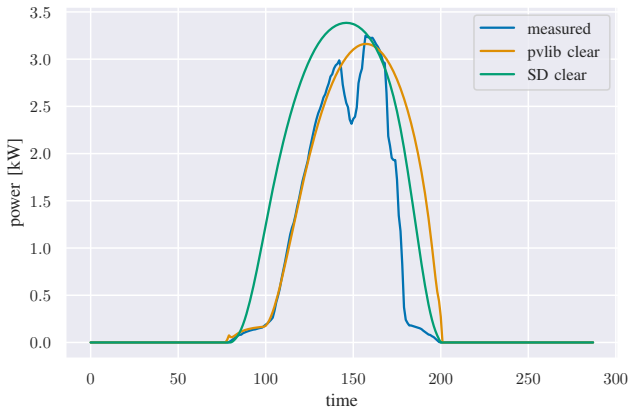}
}
\caption{A view of a single day from a system from the fleet that is oriented 
due west. Note that the green line is above the orange line.}
\label{f-west-example}
\end{figure}

Having removed the large outliers in the ``high shade'' bin, we plot the RE 
versus the system azimuth angle in figure~\ref{f-azimuth}. We observe that 
there are postive relative errors for systems that have azimuths facing east 
($90^\circ$) and west ($270^\circ$). We find that the SD model tends to over 
estimate the clear sky reference signal and therefore over estimate shade 
losses on system geometries that are due east or west. This behavior is clearly 
depicted in figure~\ref{f-west-example}. In general, we find that the proposed 
methodology works quite well for systems that are oriented roughly between 
southwest and southeast directions. Improving the accuracy on systems that are 
outside that azimuth range will be an area of future research.


\section{Conclusions}

We present an application of a novel signal decomposition framework that is 
effective at analyzing shade losses in unlabeled PV data. 
%While not nearly as  accurate as other proposed methods that make use of 
%outside information, 
This approach provides a tractable solution to estimating shade losses when no 
other 
information is available besides a power generation signal. While obviously 
hindered by not having access to e.g., reference systems or IV curves, our 
method is the first that is able to estimate shade loss to a reasonable degree 
of accuracy (usually within 10\% to 15\%) using just power data. This opens the 
door to the analysis of shade losses in fleets of distributed rooftop PV 
systems. We expect to further improve the accuracy of these methods in future 
work, refining and developing the cost functions in the SD model to better 
account for all observed conditions and orientations.

%This method has a 
%different expected accuracy than the methods based on external data such as 
%reference systems or IV curves discussed in \S\ref{s-background}, and 
%estimating shade losses to within 10 or 15\% is a significant improvement over 
%not being able to estimate them at all. In short, this 
%method makes it feasible, for the first time, to estimate the shade losses in 
%fleets of distributed rooftop PV systems. This method is well 
%suited to the application of analyzing system losses in fleets of distributed 
%PV systems, where automation and scalability are critical. Future work will 
%focus on improving results for systems with high azimuth angles references to 
%due south and on solving the problem using open source methods.

\section*{Acknowledgments}
The authors would like to thank Prof. Stephen Boyd, Luke Volpatti, Gina 
Meyers-Im, and the 
GISMo Team at SLAC National Accelerator Laboratory their for input and feedback 
on 
this work. We also recognize the Seaborn plotting package for Python, which 
made the figures possible~\cite{Waskom2021}.

	%REFERENCES
\small
\bibliographystyle{IEEEtran}
\bibliography{shade2022}

\end{document}